# CMEs in the Heliosphere: I. A Statistical Analysis of the Observational Properties of CMEs Detected in the Heliosphere from 2007 to 2017 by STEREO/HI-1

**R.A. Harrison**[1] ***(Orcid ID 0000-0002-0843-0845)***, **J.A. Davies**[1] ***(Orcid ID 0000-0001-9865-9281)***, **D. Barnes**[1], **J.P. Byrne**[1], **C.H. Perry**[1], **V. Bothmer**[2] ***(Orcid ID 0000-0001-7489-5968)***, **J.P. Eastwood**[3] ***(Orcid ID 0000-0003-4733-8319)***, **P.T. Gallagher**[4] ***(Orcid ID 0000-0001-9745-0400)***, **E.K.J. Kilpua**[5] ***(Orcid ID 0000-0002-4489-8073)***, **C. Möstl**[6] ***(Orcid ID 0000-0001-6868-4152)***, **L. Rodriguez**[7] **(Orcid ID** 0000-0002-6097-374X**)**, **A.P. Rouillard**[8], **D. Odstrčil**[9] ***(Ordid ID 0000-0001-5114-9911)***

**Abstract** We present a statistical analysis of coronal mass ejections (CMEs) imaged by the *Heliospheric Imager* (HI) instruments on board NASA's twin-spacecraft STEREO mission between April 2007 and August 2017 for STEREO-A and between April 2007 and September 2014 for STEREO-B. The analysis exploits a catalogue that was generated within the FP7 HELCATS project. Here, we focus on the observational characteristics of CMEs imaged in the heliosphere by the inner (HI-1) cameras, while following papers will present analyses of CME propagation through the entire HI fields of view. More specifically, in this paper we present distributions of the basic observational parameters – namely occurrence frequency, central position angle (PA) and PA span – derived from nearly 2000 detections of CMEs in the heliosphere by HI-1 on STEREO-A or STEREO-B from the minimum between Solar Cycles 23 and 24 to the maximum of Cycle 24; STEREO-A analysis includes a further 158 CME detections from the descending phase of Cycle 24, by which time communication with STEREO-B had been lost. We compare heliospheric CME characteristics with properties of CMEs observed at coronal altitudes, and with sunspot number. As expected, heliospheric CME rates correlate with sunspot number, and are not inconsistent with coronal rates once instrumental factors/differences in cataloguing philosophy are considered. As well as being more abundant, heliospheric CMEs, like their coronal counterparts, tend to be wider during solar maximum. Our results confirm previous coronagraph analyses suggesting that CME launch sites don't simply migrate to higher latitudes with increasing solar activity. At solar minimum, CMEs tend to be launched from equatorial latitudes while, at maximum, CMEs appear to be launched over a much wider latitude range; this has implications for understanding the CME/solar source association. Our analysis provides some supporting evidence for the systematic dragging of CMEs to lower latitude as they propagate outwards.

**Keywords** Coronal mass ejections · Heliosphere · Space weather

Corresponding author: R.A. Harrison
Richard.Harrison@stfc.ac.uk

1. STFC-RAL Space, Rutherford Appleton Laboratory, Harwell Campus, Didcot, OX11 0QX, UK
2. Institute for Astrophysics, Georg-August-University of Göttingen, 37077 Göttingen, Germany
3. Blackett Laboratory, Imperial College, London, SW7 2AZ, UK
4. Trinity College Dublin, College Green, Dublin 2, Ireland
5. Department of Physics, PO Box 64, 00014 University of Helsinki, Helsinki, Finland
6. Institute of Physics, Universitatsplatz 5, 8010 Graz, Austria
7. Royal Observatory of Belgium, Ringlaan 3, 1180 Brussels, Belgium
8. Institut de Recherche en Astrophysique et Planetologie, 9 Ave du Colonel Roche, 31028 Toulouse Cedex 4, France
9. School of Physics, Astronomy and Computational Sciences, George Mason University, Fairfax, VA 22030-4444, USA

## 1. Introduction

The results of numerous statistical studies of the coronal properties of coronal mass ejections (CMEs) have appeared in the literature since their discovery in coronagraph imagery of the early 1970s. The most comprehensive, and hence arguably the most definitive, of these studies exploit the near-continuous 20-year set of visible-light coronal observations made by the *Large Angle Spectrometric Coronagraph* (LASCO; Brueckner *et al.*, 1995) on the joint ESA/NASA *Solar and Heliospheric Observatory* (SOHO), orbiting the first Sun-Earth Lagrange (L1) point. Notable examples of such works include those by St Cyr *et al.* (2000), Yashiro *et al*. (2004) and Gopalswamy *et al*. (2009), the last of which considers over 10,000 CMEs from the so-called CDAW CME catalogue. We must also mention the more recent cataloguing endeavours undertaken by, for example, Bosman *et al.* (2012) and Vourlidas *et al.* (2017), based on analysis of the imagery from the visible-light COR-2 coronagraphs on board the twin spacecraft of the *Solar Terrestrial Relations Observatory* (STEREO; Howard *et al.,* 2008); data from the visible-light *Heliospheric Imagers*, also on STEREO, form the basis of the current study. For an overview of the results of a number of statistical analyses of CMEs ─ based on coronal observations ranging from 1971, with *Orbiting Solar Observatory*-7 (OSO-7), through to the present day ─ the reader is directed to the review of Webb and Howard (2012); we briefly discuss some the more salient of these studies below. In particular, we focus on the results of analyses of CMEs observed in the corona over the ascending phase of the solar cycle, as these are most directly comparable to the bulk of the analysis presented in the current paper.

Yashiro *et al.* (2004) presented the statistical properties of almost 7000 CMEs from the CDAW catalogue, detected during the ascending phase of Solar Cycle 23. Over that period, which extended from solar minimum in 1996 to solar maximum in 2002, the authors recorded a near monotonic increase in the occurrence of CMEs from around 200 to over 1650 per year. This increase in CME occurrence was initially accompanied by an increase in the median CME position angle (PA) span, from $43^{o}$ in 1996 to a peak value of $58^{o}$ in 1999, after which it reduced to $49^{o}$ in 2002. The authors also noted that, at solar minimum, CMEs tended to be centred at low latitudes, with, for example, the central PA of 80% of the CMEs in 1996 lying between S$24^{o}$ and N$20^{o}$. By 2002, the maximum of Cycle 23, the latitude range encompassing the central PA of 80% of CMEs had widened to S$59^{o}$ - N$51^{o}$. Unlike the case for the median PA span, the trend over the intervening years was not simply a continuously widening one. Although of less relevance to the results presented in this paper, it is worth pointing out that Yashiro *et al.* (2004) identified a clear increase in the average speed of CMEs observed in the corona over the ascending phase of Solar Cycle 23; a statistical analysis of the kinematic properties of CMEs detected in the heliosphere will be the subject of a follow-on to this paper. It is also worth noting that the results of Yashiro *et al.* (2004) are consistent with other studies of coronal CMEs, such as that of Hundhausen *et al.* (1984) based on an extensive set of observations from *Solar Maximum Mission* (SMM).

Statistical analyses such as these provide crucial information on the mechanism of mass loss from our Sun, including the onset of the ejection process. In view of this, Wang *et al.* (2011) studied the source locations of over 1000 coronal CMEs, combining LASCO data taken during the period from 1997 to 1998 ─ near the start of Solar Cycle 23 ─ with extreme-UV (EUV) disc imaging. In attempting to identify the EUV source regions of the 1078 CMEs included in their LASCO event list during those years (extracted from the CDAW catalogue), the authors found that i) for 231 of the CMEs, the source regions could not be identified due to the poor quality of the data, ii) 288 of the CMEs could be associated with clearly identifiable front-side source regions, iii) 234 had identifiable signatures over the limb, and iv) for 325 CMES, there was no clear association with eruptive structures in the EUV data.

The Wang *et al.* (2011) paper forms part of a long line of publications that, together, stress the lack of clarity of the CME onset process in terms of solar surface associations (see *e.g.* Howard and Harrison, 2013). The observability of a CME in visible light, due to the Thomson-scattering of photospheric light off free electrons confined to the ascending structures, is a function of the electron density and, of course, the Thomson scattering geometry. In contrast, the signature of a coronal process in EUV is due to line emission from a specific ion and is a function of both temperature and (the square of the) density. Thus, the direct attribution of an EUV source region to a CME is far from straightforward; some studies suggest that, rather than underlying the core of a CME, some EUV signatures ─ such as flares and dimmings ─ can be associated with CME source region footpoints (e.g. Harrison *et al.,* 1986, 2012). Moreover, as demonstrated statistically by Wang *et al*. (2011), a CME detected in coronagraph imagery may have no corresponding signature at the lower altitudes that can be observed in EUV (see also, *e.g.* Robbrecht *et al*., 2009; Kilpua *et al*., 2014).

Prior to the work of Wang *et al.* (2011), Cremades and Bothmer (2004) had also undertaken a detailed CME source region study. Those authors concluded, in particular, that CMEs were more likely to be associated with active and decaying regions. For those CMEs with an identified source region, both Cremades and Bothmer (2004) and Wang *et al.* (2011) compared, statistically, the apparent PA of the source region to that of the overlying CME, concluding that most CMEs are apparently deflected towards the equator ─ at least near solar minimum ─ as they propagate outwards from the Sun. Whereas Cremades and Bothmer (2004) and Wang *et al*. (2011) discuss equatorward drag in relation to potential CME source regions, there are a number of studies of equatorward drag based on coronagraph observations alone, with no reference to CME source regions; this is discussed later.

It is also worth noting here some results of a study undertaken by Michalek and Yashiro (2013), who investigated the relationship between CMEs and active regions, through the analysis of almost 700 CMEs observed near the peak of Solar Cycle 23 (from 2001 to 2004). The authors concluded that CMEs are more likely to be associated with mature active regions with complex magnetic fields. Indeed, they claim that the fastest CMEs are associated with active regions that exhibit extreme magnetic complexity. In contrast, they demonstrated that wider CMEs tend to originate from magnetically-simple source regions. Clearly, longer-term statistical studies of such parameters as CME propagation direction and angular width would enable a better understanding of such results, particularly in terms of their solar cycle variation.

Many of the aforementioned works document statistical analyses of the morphology and kinematic properties of CMEs based on visible-light coronagraph observations. These works have provided important information on the physics of CME onsets and the early propagation processes. However, for most events, after the CME had been observed to pass through the corona, its only subsequent detection was as a result of its passage over one of the relatively sparse set of space-based *in situ* solar wind observatories. CMEs have been detected *in situ* near Earth by, for example, such spacecraft as SOHO (Domingo *et al.*, 1995), Wind (Acuna *et al*., 1995) and *Advanced Composition Explorer* (ACE; Stone *et al.*, 1998); they have, moreover, been detected at locations distant from Earth, by spacecraft situated near other planets, such as Mars and Venus, and elsewhere in the heliosphere (see *e.g.* Richardson, 2014). As a result, over the majority of time since their initial discovery, observation of CMEs has been restricted to their initial phase in the corona and their subsequent, and fortuitous, *in situ* detection, often hundreds of solar radii from the Sun. The advent of wide-angle, visible-light heliospheric imaging - by the ground-breaking *Solar Mass Ejection Imager* (SMEI; Eyles *et al.,* 2003) instrument flown on board the low Earth-orbiting *Coriolis* mission and, subsequently, by the flagship *Heliospheric Imager* (HI; Eyles *et al.,* 2009) instruments on NASA's STEREO mission - has extended the routine imaging of CMEs well beyond the 7.5$^{o}$ elongation outer limit of the SOHO/LASCO C3 field of view. Exploitation of heliospheric imagery from STEREO/HI, in particular, has demonstrated that CMEs can be tracked out to 1 AU and beyond, and provided evidence that CME kinematic properties and morphology can evolve significantly throughout their propagation. This evolution can be due to their interaction with the background solar wind - including Stream Interaction Regions (SIRs)/Co-rotating Interaction Regions (CIRs) and fast solar wind streams - as well as other CMEs (see *e.g.* Byrne *et al*., 2010; Davis *et al.,* 2010; Savani *et al*., 2010; Vršnak *et al.,* 2010; Harrison *et al*., 2012; Lugaz *et al*., 2012; Liu *et al.*, 2013; Mishra and Srivastava, 2013; Maričić *et al.,* 2014; Rollett *et al.,* 2014; Temmer *et al.,* 2014; Wang *et al.,* 2014; Zhao *et al.,* 2016).

By necessity, CME catalogues derived from single-spacecraft coronagraph imagery cite event speeds in the plane of the sky. In contrast, the availability of wide-angle imaging even from a single vantage point allows us to explore the 3D nature of CME propagation through the heliosphere; the extended elongation range of the data enables us to derive such 3D information through geometrical modelling. More specifically, fitting the time-elongation profiles of features of outward-propagating CMEs over large distances from the Sun (*e.g.* Davies *et al.*, 2009) allows us, in principle, to determine the latitude and longitude (and radial speed) of propagation of each CME. Thus, we can determine whether or not a particular event might be directed towards Earth, Venus, Mars or indeed any other solar system "target" (see Möstl *et al.,* 2012, 2017).

It is the purpose of this paper to focus on CMEs in the heliosphere by providing a benchmark for the cataloguing and statistical analysis of such events, which allows us to address uniquely a range of issues pertaining to CME propagation and evolution, and a number of concepts relating to CME onset models. It naturally sits alongside the established coronal CME catalogues that are mentioned in this paper, thereby setting the stage for a comprehensive analysis of the relationship between CMEs in the corona and heliosphere that should confirm or refute the findings of studies such as those mentioned above. This should pave the way for a far more complete view of CMEs, their sources, evolution and propagation, and by combining with "ground-truth" data from *in situ* measurement at specified locations, we are also much more able to assess the impacts of events that have been imaged in the corona and heliosphere.

## 2. STEREO/HI and the HELCATS Project

The twin spacecraft of the STEREO mission were launched in October 2006 into heliocentric ecliptic orbits, one (STEREO-A) leading and the other (STEREO-B) lagging the Earth in its orbit (Kaiser *et al.,* 2007; Driesman *et al.,* 2007). The radial distances of the two spacecraft from the Sun are such that each moves relative to the Sun-Earth line by some 22.5° per year. The dual off Sun-Earth line vantage points afforded by the STEREO mission has enabled a number of unique capabilities such as 3D reconstruction of low coronal features, and stereoscopic imaging of solar transients propagating along the Sun-Earth line. After some eight years, the STEREO spacecraft entered superior conjunction, going behind the Sun with respect to the Earth. The STEREO-A spacecraft entered superior conjunction in March 2015 and emerged in full health in July 2015; at the time of writing, STEREO-A continues to operate nominally. Unfortunately, contact with STEREO-B was lost prior to the spacecraft entering superior conjunction. It should be noted that, for several months either side of superior conjunction, STEREO-A was operated in a reduced telemetry mode, necessitated by unforeseen heating of the high gain antenna; the mitigation strategy involved use of antenna off-pointing. In fact, it was during testing of such reduced operations for STEREO-B (in October 2014) that contact with that spacecraft was lost; at the time of writing, endeavours to re-establish contact with STEREO-B are continuing.

The imaging capabilities of the STEREO mission are provided by the *Sun Earth Connection Coronal and Heliospheric Investigation* (SECCHI; Howard *et al.,* 2008) package. SECCHI provides disk imagery at four EUV wavelengths (with EUVI, the *Extreme Ultra Violet Imager*), as well as visible-light imagery of both the corona and the heliosphere, the latter out to 1 AU and beyond, which is achieved by a combination of two coronagraphs (COR-1 and COR-2) and the *Heliospheric Imager* (HI; Eyles *et al.,* 2009; Harrison *et al.,* 2008); imagery from the latter forms the basis for the work presented here. The HI instrument on each STEREO spacecraft comprises two wide-angle, visible-light cameras: the 20° diameter field of view (inner) HI-1 camera ─ the bore-site of which is nominally aligned at 14° elongation (angle from Sun-centre) in the ecliptic plane ─ and the 70° diameter field of view (outer) HI-2 camera ─ the bore-site of which is nominally aligned at 53.7° elongation in the ecliptic plane. Prior to superior conjunction, HI on STEREO-A imaged to the (solar) east of the Sun-spacecraft line and HI on STEREO-B, to the west; since emerging from conjunction, for STEREO-A, this is now reversed. The HI-1 and HI-2 cameras are each based on 2k × 2k CCD detector systems, although, for routine science usage, images are binned to 1k × 1k prior to downlink. Despite a reduction in overall spacecraft telemetry since launch ─ imposed by increasing distance from Earth ─ the cadences of the STEREO/HI-1 and HI-2 science imagery have remained at their nominal values of 40 min and 120 min, respectively, practically throughout the mission to date, highlighting the unique nature of these observations.

The STEREO/HI instruments allow the imaging of plasma density enhancements such as CMEs in the heliosphere through the detection of Thomson-scattered photospheric (visible) light. STEREO/HI has provided ground-breaking observations, not only of CMEs (the prime scientific objective of the STEREO mission) but also of a variety of other phenomena, both inside the heliosphere and beyond (e.g. Harrison *et al.* 2009).

Figure 1 presents STEREO/HI-1 images of six different CMEs detected in the $20^{o} \times 20^{o}$ field of view of either the HI-1 camera on STEREO-A (HI-1A; top two rows) or the HI-1 camera on STEREO-B (HI-1B; bottom two rows). The events were chosen at random, one from each of the "good", "fair" or "poor" CME categories, from the HELCATS (*Heliospheric Cataloguing, Analysis and Techniques Service*) manual CME catalogue (HICAT), which we discuss in more detail later. Each image is shown in a background-subtracted format (red colour table; upper panel of each pair) and in a difference-image format (grey scale; lower panel of each pair). For the former, a daily background (comprising mainly F-corona) is subtracted, whilst, for the latter, the image is displayed with the previous image subtracted. These two formats are used to reveal different aspects of CME topology and evolution. Difference images are excellent for highlighting variability (the white/black structures identify regions of excess/depleted density relative to the previous frame) whilst background-subtracted images show the more fundamental structure of CMEs more effectively. As mentioned above, the HELCATS manual cataloguing procedure identifies CMEs as good, fair or poor, depending on the clarity of the CME structure (the definition of these categories will be discussed in more detail below); the left, middle and right hand columns of Figure 1 show examples of good, fair and poor events, respectively.

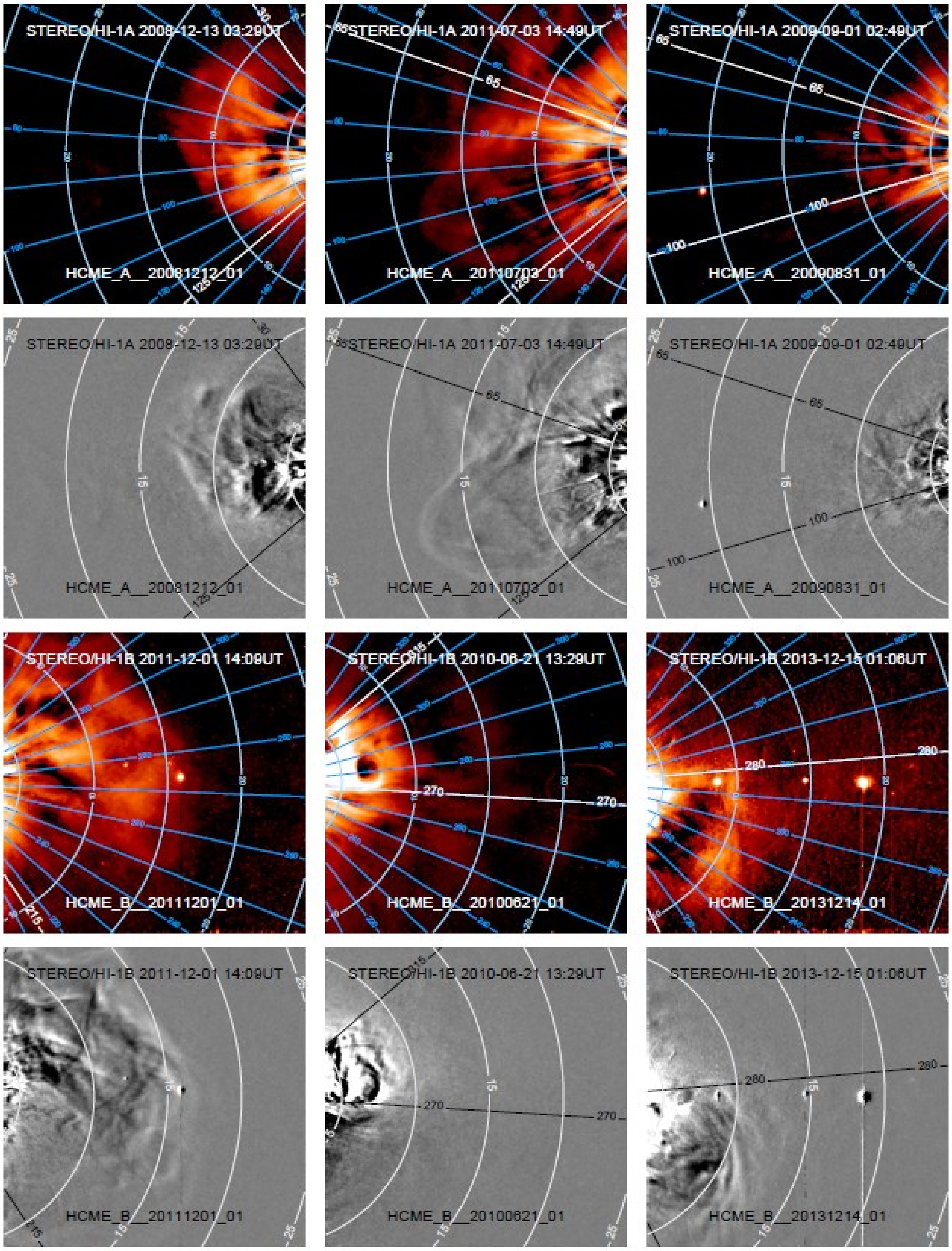

**Figure 1** Images of six heliospheric CMEs detected by the STEREO/HI-1 cameras. Observations of each CME are shown in both background-subtracted format (top frame of each pair; F-coronal/residual stray-light signal is removed) and difference-image format (bottom frame of each pair; previous frame is subtracted). The top three events were detected by HI-1A and the bottom three, by HI-1B. From left to right, the columns contain CMEs categorised in the HELCATS manual (HICAT) catalogue as good, fair and poor. The unique HICAT event ID for the corresponding CME is given at the foot of each image.

The STEREO/HI-1 manual CME catalogue (HICAT), analysis of which forms the basis of this paper, was generated under the auspices of the HELCATS project (which ran from May 2014 to April 2017), funded under the European Union's Seventh Framework Programme for Research and Technological Development (EU FP7). The RAL Space-led HELCATS consortium, which included 8 beneficiaries from 7 European countries plus two third party collaborators (one from the US), capitalised on long-established European expertise in heliospheric imaging, particularly its lead of STEREO/HI, whilst also exploiting Europe's experience in solar and coronal imaging and the interpretation of *in situ* and radio measurements of the solar wind. The general aims of HELCATS involved: i) cataloguing of both transient (CME) and background (SIR/CIR) solar wind structures imaged by the STEREO/HI instruments, including estimates of their kinematic properties derived through the application of a variety of both established and speculative modelling techniques; ii) verification of the derived kinematic properties of the transient and background solar wind components, such that the validity of the modelling techniques could be assessed, by comparing both with solar source observations and with *in situ* measurements throughout the inner heliosphere; iii) evaluation of the potential for using these derived kinematic properties to also initialise advanced numerical models; iv) assessment of the complementarity of heliospheric imagery and radio-based methods to detect structures and diagnose processes in the inner heliosphere (specifically Type II burst and Interplanetary Scintillation, IPS, data); and v) provision of straightforward access to the HELCATS products and methodologies, thereby enabling heliospheric imaging observations to be understood and exploited more widely. Further information, and access to the HELCATS products, is available from the HELCATS website at https://www.helcats-fp7.eu.

The main task of Work Package 2 (WP2) of the HELCATS project involved the generation of a catalogue of CMEs detected in the heliosphere through visual inspection of background-subtracted and difference STEREO/HI-1 images; this manual STEREO/HI-1 CME catalogue is referred to as HICAT. From the HELCATS website (https://www.helcats-fp7.eu), HICAT can be accessed via the PRODUCTS tab by selecting “WP2 : HICAT” (or directly via the link https://www.helcats-fp7.eu/catalogues/wp2_cat.html). HICAT includes the basic observational parameters of each identified CME, namely:

i) The date and time (in UTC) of its first definitive observation in the HI-1 imagery (to an accuracy defined by the image cadence of 40 min, notwithstanding limitations in data coverage);
ii) The observing spacecraft (A or B);
iii) Its northernmost PA extent (rounded to the nearest 5°), including an indicator as to whether the CME appears to extend beyond the northernmost extent of the HI-1 field of view;
iv) Its southernmost PA extent (rounded to the nearest 5°), including an indicator as to whether the CME appears to extend beyond the southernmost extent of the HI-1 field of view;
v) A measure of the confidence that the eruption is, by definition, a CME, categorised (albeit somewhat subjectively given the nature of manual cataloguing endeavours) as either good, fair or poor.
vi) A unique identifier for each CME, constructed from a combination of some of the aforementioned parameters.

It is important to note that a threshold in terms of the minimum PA extent is applied in the identification of CMEs; this threshold is set at 20° to avoid the cataloguing of the numerous blob-like features that are detected by STEREO/HI in the solar wind. It should also be borne in mind that the latitudinal extent to which this PA extent corresponds depends upon a CME’s 3D propagation direction. The accuracy of $5^{o}$ to which the northern and southernmost PA extents of a CME are quoted reflects the difficulty in assigning definitive values to these parameters due to the diffuse and variable nature of CMEs and their boundaries.

As noted above, HICAT CMEs are ascribed a category according to their clarity, being assigned as good, fair or poor; we expand on that categorisation here. As with coronal CME catalogues, such a classification is extremely difficult to quantify precisely and is therefore highly subjective; hence these assignments should be considered only as a guide by catalogue users. However a good event in HICAT is one that we believe any experienced user would unambiguously consider as a CME; conversely, a poor event is one that, while our cataloguer considers it likely to be a CME, is acknowledged as being sufficiently ambiguous (in terms of such factors as brightness, topology, or even due to the presence of data gaps) that some STEREO/HI users may dispute that assertion. While information on these poor events is available for download via the HELCATS website, it is not necessarily recommended that this information be used for individual CME analyses without further critical appraisal. The fair events are those that we consider as falling between the two aforementioned categories, relatively clear but not indisputable.

Release notes describing the generation of the catalogue, and the parameters therein, can also be accessed via https://www.helcats-fp7.eu/catalogues/wp2_cat.html. For ease of future exploitation, the HELCATS catalogues, including the manual HICAT CME catalogue, are under version control and given a DOI (10.6084/m9.figshare.5803152). At the time of writing, Version 5 (released 2018-01-19) of HICAT is installed on the HELCATS website.

Version 5 of the HICAT catalogue was derived from HI-1A images that were taken between 1 April 2007 (the start of the science phase of the STEREO mission) and 18 August 2014 (the last date for which nominal cadence imagery is available prior to the start of reduced operations leading up to superior conjunction); this version of the catalogue also includes CMEs identified in post-superior conjunction HI-1A imagery, taken after 17 November 2015 when full operations recommenced. For STEREO-B, this version of HICAT includes CMEs identified in HI-1B images taken between 1 April 2007 and 27 September 2014 (the latter corresponding to the last date for which images were available prior to loss of contact with the spacecraft). At the time of writing, HICAT is still being updated (despite the fact that the HELCATS project has finished) to reflect the fact that HI on STEREO-A continues to operate successfully; should full contact be re-established with STEREO-B ─ with its loss having had no detrimental effect on the HI instrument ─ HICAT will be updated accordingly.

The current study includes analysis of CMEs up to the end of August 2017. Version 5 of HICAT lists 2059 manual heliospheric CME identifications in the STEREO/HI-1 data up to that date, with 1901 of those occurring prior to the solar conjunction phase. Of those 2059 entries, 1630 are listed as either good or fair events, 898 for HI-1A and 732 for HI-1B, and the rest (429) as poor events.

Given the characteristics of the HI instruments and the nature of the STEREO orbit, the HICAT manual cataloguing is undertaken independently for the two spacecraft. Many of the HICAT CMEs will have been detected both from the vantage point of STEREO-A and that of STEREO-B; this issue will be discussed briefly below but is returned to in more detail in a later study.

We note that in WP3 of the HELCATS project, the majority of the HICAT catalogue entries for those CMEs identified as being good or fair are augmented with estimates of the CME's kinematic properties ─ more specifically launch time, 3D propagation direction, and radial speed ─ derived from application of the Fixed Phi Fitting (FPF), Harmonic Mean Fitting (HMF), and Self-Similar Expansion Fitting (SSEF) techniques (see Davies *et al.*, 2012 and references therein) to their time-elongation profiles manually extracted from combined HI-1/HI-2 time-elongation maps (J-maps); this endeavour is briefly discussed by Möstl *et al.* (2017) but a more detailed description will be given in a subsequent paper (Barnes *et al.,* 2018). This extended version of the HICAT catalogue, known as the HIGeoCAT (because of its use of geometrical models; see Möstl *et al.*, 2017), can be accessed via https://www.helcats-fp7.eu/catalogues/wp3_cat.html. For those CMEs identified as being observed in HI imagery from both spacecraft, the analogous stereoscopic techniques (see Davies *et al.*, 2013 and references therein) have also been applied.

## 3. Observations

Panels a and b of Figure 2 present occurrence frequencies of the heliospheric CMEs in the HICAT catalogue that were detected by HI-1A (panel a) and HI-1B (panel b), from the start of April 2007 to the end of August 2017. The data are binned by calendar month, and then normalised to CMEs per day. Obviously the STEREO-B data only extends until the end of September 2014, when contact with the spacecraft was lost. The gap in STEREO-A data, covering a portion of the years 2014 and 2015, corresponds to superior conjunction itself and the surrounding period of reduced operations. CMEs are colour coded according to their quality, i.e. be they good, fair or poor (the total height of each bar indicates the total rate of CMEs). In each plot, the total rate of CMEs detected by the other spacecraft is indicated for ease of comparison (as diamonds). It should be noted, and is justified later on, that there has been no "duty cycle" correction applied to the data (*i.e.* no correction for gaps in continuous operation). Unsurprisingly, both HI-1A and HI-1B CME occurrence profiles are generally consistent with the expected variation with solar cycle. During the solar minimum years of 2008 and 2009, the heliospheric CME rate for HI-1A was of order 0.05-0.1 per day for events classified as good, and typically 0.15-0.2 per day in total. The corresponding rates are marginally lower for HI-1B. The difference in CME rates between HI-1A and HI-1B may be due, in part, to the fact that the HI-1B camera (unlike HI-1A) sometimes undergoes a jitter that appears to be associated with an interplay between spacecraft reaction wheel speed and the mounting of the camera focal plane assembly (see Tappin, 2017; Tappin, Eyles and Davies, 2017). This leads to more difficulty in extracting the background (this has implications even in the generation of difference images). In practice, this means that the HI-1A instrument has a slightly lower effective threshold than HI-1B for the detection of weak CMEs.

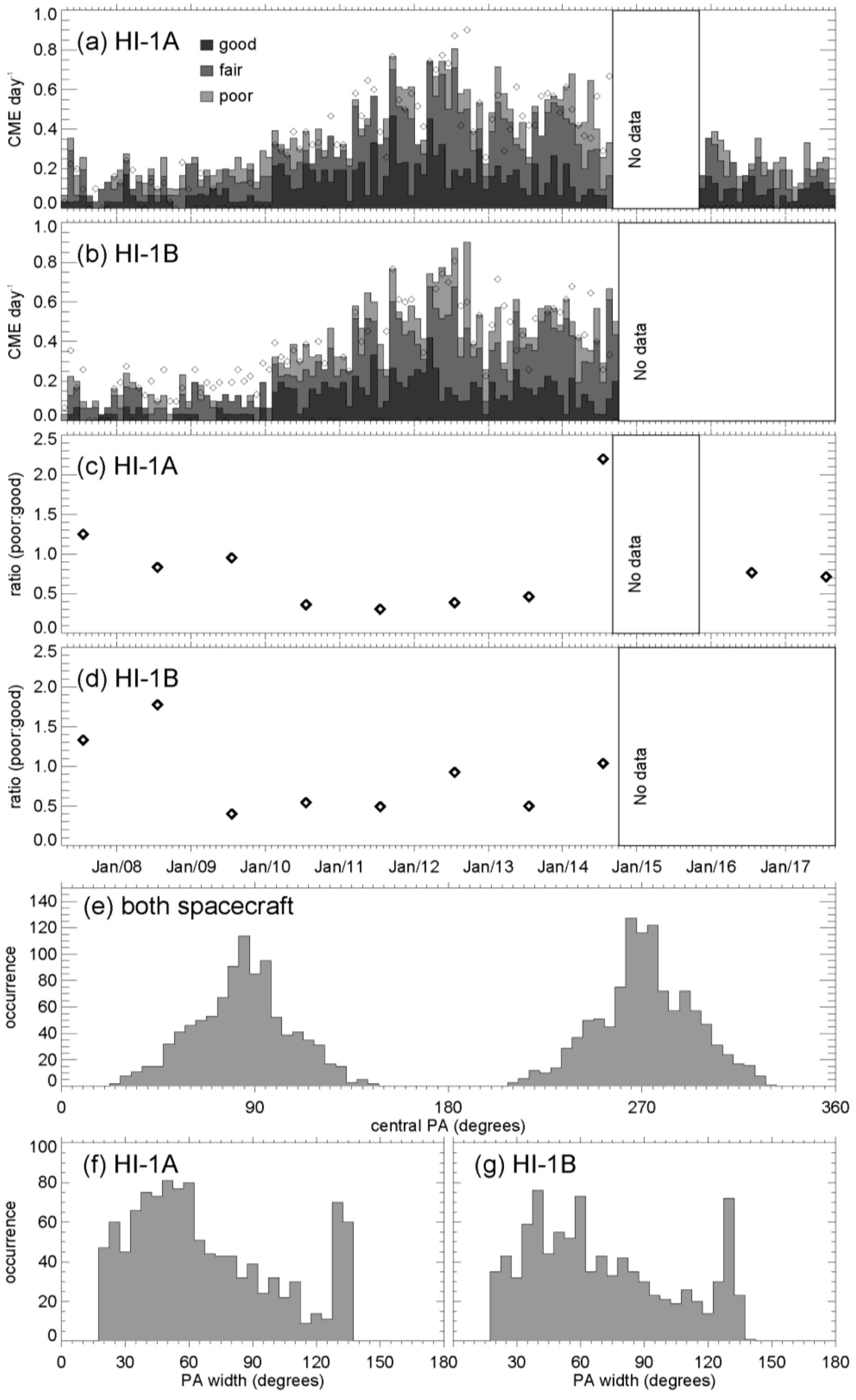

**Figure 2** Histogram of daily-average rates (binned by calendar month) of HICAT CMEs, identified in HI-1A (panel a) and HI-1B (panel b) images. Each bar is sub-divided according to CME quality, into good, fair and poor events. The diamonds represent the total rate of HICAT CMEs imaged by the corresponding camera on the other spacecraft. Panels c and d present, for HI-1A and HI-1B, respectively, the yearly-averaged ratio of poor to good events. Panel e shows the frequency distribution of CME central PA for all HI-1A and HI-1B detections (including good, fair and poor events), with a bin-width of 10$^{o}$. Panels f and g present (separately) the HI-1A and HI-1B CME PA width distributions, respectively, for all identified CMEs, again calculated with a bin-width of 10$^{o}$.

In fact, the heliospheric CME rate varies little from the start of the STEREO science observations, in April 2007, up to January 2010. Indeed, prior to January 2010, there is little indication that the rate will rise at all. An increase in the CME rate in January 2010, to a value of order 0.2 per day for good HI-1A events (≈0.3-0.4 per day for all events), is followed by a second increase in February-March 2011 that leads into a period of peak CME activity lasting for around two years. The maximum in CME activity, in early to mid-2012, corresponds to a rate of good HI-1A CME events of around 0.3 per day, with a total CME rate of about 0.7-0.8 per day. The HI-1B CME rate follows the same general trend. However, while (as mentioned above) the HI-1B CME rate tends to be consistently lower than that for HI-1A during the solar minimum years, this is not the case during periods of higher solar activity. The decline in the rate of heliospheric CMEs identified in post-conjunction HI-1A images from 2015 to 2017, with values similar to those witnessed in early 2010, continues the gradual reduction in CME rates over the descending phase of Solar Cycle 24 observed by both spacecraft after the 2012 maximum, heading towards an anticipated minimum in 2018-2019.

One may speculate on a potential variation in the observability of CMEs over the solar cycle, in that events during solar minimum may tend to be less intense (*i.e.* less dense), and hence more amorphous, than those at solar maximum. Whilst such a variation is not immediately obvious from Figures 2a and 2b, this is explored further. To this end, Figures 2c and 2d show, for HI-1A and HI-1B respectively, the yearly-averaged ratios of the number of poor CMEs to the number of good CMEs. Both plots indicate a general tendency for lower ratios during periods of higher solar activity, suggesting that, when the Sun is more active, CMEs tend to be clearer and brighter. However, the uncertainties in the plotted values, in terms of counting statistics based upon the monthly CME rates, range from 20%, when the Sun is active, to values as high as 60% when activity is low. So, although tenuous, in a statistical sense at least, it is tempting to speculate on the basis of these observations that solar maximum yields clearer, brighter CMEs in the heliosphere and, conversely, that solar minimum CMEs to be to fainter and less clear.

Note that the STEREO spacecraft are not stationary in the Sun-Earth reference frame, in that the Earth-Sun-spacecraft angles vary considerably over the period under investigation (by ≈230$^{o}$ for STEREO-A and by ≈170$^{o}$ for STEREO-B). However we would not expect the rate of CMEs, as observed from a single spacecraft, to be dependent on longitude. In contrast, the angular (specifically longitudinal) separation between STEREO-A and STEREO-B varied considerably over the period during which both spacecraft were operating nominally, so the proportion of CMEs that are observed by both HI-1A and HI-1B varies significantly. This will be discussed later.

Figure 2e presents the distribution of the central PA for all HICAT CME events (*i.e.* poor, fair and good), identified in HI-1, where angles of 0$^{o}$, 90$^{o}$, 180$^{o}$ and 270$^{o}$, represent solar north, east, south and west, respectively. The central PA is calculated as the PA that is midway between the northern and southernmost PAs cited in HICAT. The distribution exhibits two peaks, both of which are centred near the solar equator (at around 90$^{o}$ and 270$^{o}$ PA), illustrating that well-known fact that CMEs tend to be equatorial phenomena (*e.g.* Webb and Howard, 2012). The twin peaks of the distribution are relatively wide, spreading over PA ranges of 20$^{o}$ - 150$^{o}$ and 210$^{o}$ - 330$^{o}$. The events detected by HI-1A and HI-1B prior to superior conjunction are incorporated into the peaks centred near 90$^{o}$ and 270$^{o}$, respectively. Due to the 180$^{o}$ rotation of the STEREO-A spacecraft as it emerged from superior conjunction, to maintain communication with Earth, HI-1A events detected since conjunction are incorporated in the peak centred at around 270$^{o}$. The two peaks of the distribution are similar, except that the east limb (90$^{o}$-centred) peak exhibits a slight asymmetry around 90$^{o}$, appearing to slightly favour identification of events north of the equator. This distribution is, of course, accumulated over an extended period and its variation as a function of the solar cycle is examined below.

Figures 2f and 2g show the PA width distributions for all HI-1A and HI-1B-detected CMEs (*i.e.* good, fair and poor), respectively. Events that extend either northward or southward (or indeed both) beyond the limits of the HI-1 field of view result in spikes in the distributions near 130$^{o}$ to which their true PA extents are artificially truncated. The distributions demonstrate that CMEs exhibit a wide range of PA widths, upward from the 20$^{o}$ lower limit imposed in the cataloguing, peaking between 40$^{o}$ and 60$^{o}$. Again, the nature of this distribution as a function of the solar cycle is discussed below.

The distributions presented in Figure 2 provide a unique benchmark of CME activity in the heliosphere over a period extending from solar minimum to solar maximum, and beyond.

### 3.1 CME Rates and the Solar Cycle

Extrapolating the heliospheric daily CME rates presented in Figures 2a and 2b (and quoted earlier in the text) to CME rates for the whole heliosphere needs to take into account the fact that the HI-1 instruments do not image over the entire 360$^o$ of PA that is, conventionally, covered by coronagraph fields of view. Of course multiplying the rates by the ratio between 360$^o$ and the HI-1 PA extent is inappropriate since CME emergence is not equally distributed in latitude, with CMEs tending to launch from the equatorial region, as evidenced by Figure 2 (see *e.g.* Webb and Howard, 2012 and references therein). The HI-1 (and HI-2) fields of view are centred on the ecliptic plane, in keeping with the STEREO mission focus on Earth-directed events. The opening angle of the HI-1 field of view from Sun-centre is some 140$^o$ in PA, given its 20$^o$ lateral extent and inner elongation limit of around 4$^o$, although its "square" shape means that the PA coverage reduces with increasing elongation. Considering the distribution of the central PA and PA width (relating to central latitude and latitudinal width) of CMEs observed in coronagraph imagery ─ as presented by such authors as Webb and Howard (2012) ─ we can surmise that, given the orientation and extent of its field of view, HI-1 will actually detect most CMEs even at solar maximum when they tend to be both wider and emerge over a larger range of latitudes (see later discussion). That said, we note again that for HI-1B images, due to the instrument jitter mentioned above (Tappin, 2017; Tappin, Eyles and Davies, 2017), the threshold for CME detection is effectively slightly lower than for HI-1A. We need also to bear in mind potential Thomson scatter geometry sensitivity effects whereby CMEs well out of the plane of the sky of the HI cameras are inherently less bright (*e.g.* Howard and Tappin, 2009); conversely, selection effects may arise due to the fact that CMEs that are out of the plane of the sky appear wider in terms of their PA extent (only for CMEs travelling in the plane of the sky does PA extent equate to latitudinal extent) although this is, of course, also the case for CMEs imaged by coronagraphs.

As alluded to above, the gradual motion of the two STEREO spacecraft, relative to one another and Earth, means that the overlap in longitudinal coverage of the HI-1 fields of view varies significantly with time. Just after launch, the HI-1A and HI-1B fields of view encompassed completely different regions of space; when the spacecraft were 180$^o$ apart, their fields of view virtually overlapped. Despite the potential issues highlighted above, if we assume that we identify most CMEs ejected from the corresponding solar hemisphere as they pass through the field of view of one, or other, of the HI-1 instruments, we can potentially argue that, for either instrument, the heliospheric CME rate is approximately half of the total CME rate of the Sun. This suggests that the total daily rate for CMEs entering the heliosphere at solar minimum is approximately 0.3-0.4 per day for all CMEs (0.1-0.2 per day for good events) rising to 1.4-1.6 per day (0.6 per day for good events), at solar maximum. This is discussed further below.

Above we mention some of the spatial issues related to the determination of total CME rates from CME detections from either HI-1 instrument. In addition, we should consider temporal corrections, due either to duty-cycle effects or gaps in observations. The HI image cadence has remained constant ─ 40 and 120 min, for the HI-1 and HI-2 instruments, respectively ─ over the vast majority of nominal mission operations (notwithstanding reduced operations around superior conjunction, and the loss of contact with STEREO-B), despite the significant reduction in overall telemetry as the spacecraft moved further from Earth; this demonstrates the high regard in which this instrument is held. Each HI-1 image is the sum of 30 exposures, each of 40 s duration (Eyles *et al.,* 2009), with each exposure being scrubbed of cosmic ray hits (by which terminology we also include solar energetic particles) prior to summing (and subsequent binning) on board, giving an effective HI-1 duty cycle of 50%. Note that the exposure sequence is taken over a period of 30 mins (Eyles *et al.* 2009), which includes the time taken to clear the CCD prior to each exposure and read the CCD after each exposure, as well as a short amount of "dead time" between each exposure. Thus the longest time between successive images for which the instrument is not actually exposing is only 10 mins (*i.e.* the dead time between each exposure sequence). As even an extremely fast CME (> 3000 km/s) would take a minimum of 4 hours to cross the HI-1 field of view (longer, if it was propagating out of the plane of the sky), a duty-cycle correction is wholly unnecessary.

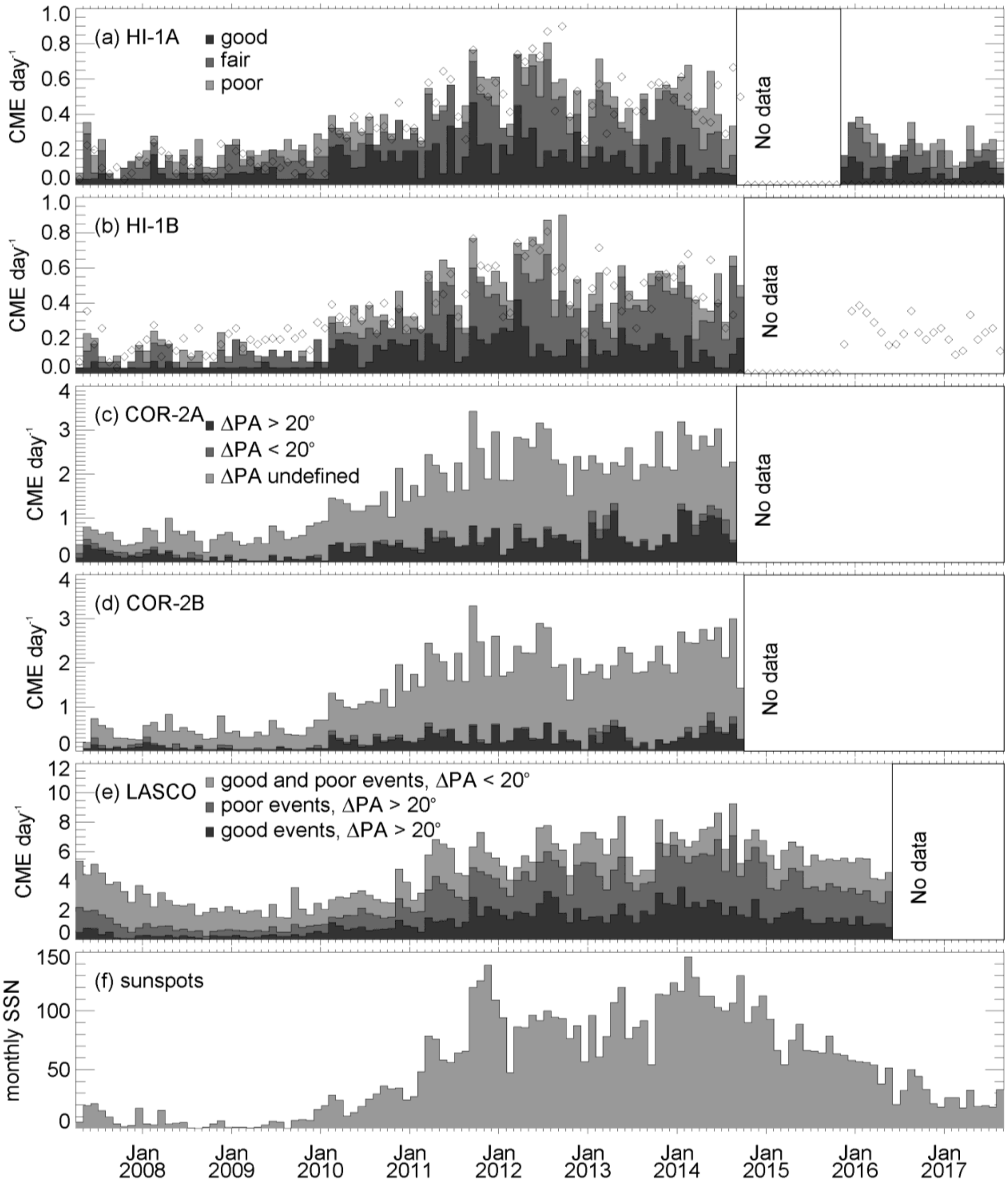

**Figure 3** Histogram of daily-average rates (binned by calendar month) of HICAT CMEs detected by (a) STEREO/HI-1A and (b) STEREO/HI-1B (reproduced from Figure 1) compared to CME rates from (c) STEREO/COR-2A and (d) STEREO/COR-2B and (e) SOHO/LASCO. The diamonds in panels a and b represent the total rate of HICAT CMEs imaged by the corresponding camera on the other spacecraft. LASCO CME rates are derived from the CDAW catalogue, and COR-2 rates, from the catalogue presented by Vourlidas *et al.* (2017). Panel f shows the monthly sunspot count over the same period, provided by the Solar Influences Data Analysis Centre (SIDC) of the Royal Observatory of Belgium.

Of more concern, however, is the potential impact of data gaps. Firstly, it is worth noting that, for the majority of months of nominal operation, data return (in terms of the number of HI-1 images) is well in excess of 95% of that expected, demonstrating the remarkable continuity of the HI imagery. However, it is more important to consider the duration of individual data gaps to ascertain if they could potentially lead to CMEs passing undetected.

A gap in the HI-1 data would need to exceed 12 hours duration to lead to the possible non-detection of a relatively fast CME (> 1000 km/s ─ faster than the majority of events), assuming that the CME could be observed across the entire field of view; this would correspond to a gap of around 20 images at nominal 40-minute cadence. Of course, to completely miss a CME even in these circumstances would require it to be launched near the start of the data gap. There are 49 gaps longer than this in the nominal-cadence HI-1A dataset (from April 2007 to August 2017, outside the conjunction phase). Only four of these gaps lasted longer than 50 images (*i.e.* over 33 hours), the longest, at 121 hours occurring just prior to the onset of the conjunction phase. The HI-1B dataset manifests 31 gaps longer than 20 images in duration, again with only four longer than 50 images (these range from 39 to 45 hours in length). It is not straightforward to assess the required correction to account for these data-gaps but, making the relatively extreme assumption that a single CME was missed for each gap exceeding 20 images, this would lead to the loss of 49 HI-1A CMEs. This amounts to an average over the observation period of 0.01 CMEs per day, making it insignificant in terms of Figure 2a and 2b.

Panels a and b of Figure 3 reproduce panels a and b of Figure 2, respectively, to facilitate comparison of the occurrence frequency of heliospheric CMEs with that of coronal CMEs observed by STEREO/COR-2A (panel c), STEREO/COR-2B (panel d) – based on the COR-2 catalogue described by Vourlidas *et al.* (2017; http://solar.jhuapl.edu/Data-Products/COR-CME-Catalog.php) – and SOHO/LASCO (panel e; derived from the CDAW catalogue https://cdaw.gsfc.nasa.gov/CME_list/). Also included is the monthly sunspot number profile (panel f) provided by the Solar Influences Data Analysis Centre (SIDC; http://sidc.ima.be) of the Royal Observatory of Belgium. Note that the end dates of the CME rates displayed for COR-2A and LASCO are dictated by the contents of the relevant catalogues, rather than a lack of available imagery. Clearly the underlying solar cycle trend is the same for sunspots, and coronal and heliospheric CMEs. However, whereas the sunspot number increases from solar minimum to maximum by approximately two orders of magnitude (Figure 3f), the total heliospheric CME rates (good, fair and poor CMEs) appear only to increase by a factor of 4 to 5 (Figures 3a and 3b).

Some of the events included in the COR-2 CME catalogue (Vourlidas *et al.* 2017), from which Figures 3c and 3d are derived, have been analysed in a semi-automatic manner to provide a specified PA width (and central PA); many others have not yet been analysed in this way (note that seemingly more COR-2A than COR-2B CMEs have been analysed thus). COR-2 CMEs in Figures 3c and 3d are colour-coded according to their PA width into those wider than 20$^o$, those narrower than 20$^o$, and those whose widths are not (yet) provided. The total COR-2 CME rates are of order 0.5 CMEs/day at solar minimum, increasing to some 3 CMEs/day at maximum. These rates are around four times greater than the corresponding rates of CMEs identified in either HI-1A or HI-1B images. This discrepancy can be accounted for by a combination of three main factors:

- arguably most significantly, accounting for a factor of two difference in CME rates between HI-1 and COR-2, the difference in the "longitudinal" field of view coverage, i.e. the fact that each HI instrument images only one side (east or west) of the Sun-spacecraft line (see discussion above related to CME rates projected to the full heliosphere);
- less significantly given that CMEs are predominantly an equatorial phenomenon, the fact that the HI-1 cameras don't image the polar regions;
- and the fact that most of the CMEs in the COR-2 catalogue are of undefined PA width, and we are comparing with heliospheric CMEs wider than 20$^o$ (although most of the COR-2 CMEs with defined widths are wider than 20$^o$, this may not be representative of the entire population).

Hence, with the view that CMEs are mostly an equatorial phenomenon and that a significant fraction of CMEs of undefined width in the COR-2 catalogue are likely to be narrower than 20$^o$, the rates of CMEs imaged by STEREO in the outer corona and heliosphere are not inconsistent. Of course, additional factors, such as relative sensitivity of the instruments, may also contribute to some degree to this discrepancy but it is difficult to ascertain at what level this would be.

The SOHO/LASCO daily CME rates (derived from the CDAW CME list and presented in Figure 3e) appear to show more significant CME activity over the declining months of Solar Cycle 23, *i.e.* over 2007, than do either STEREO/HI or COR-2. This is particularly evident when narrow events (< 20$^o$ in PA extent; light grey) are also taken into consideration; when considering only events wider than 20$^o$ (dark and medium grey), this effect is much less pronounced. The proportion of narrow events in the CDAW catalogue is seemingly much higher during periods of low solar activity; we suggest that these may not be conventional CMEs, but, instead, smaller-scale blobs originating from coronal streamers (*e.g.* Sheeley *et al.*, 2009; Rouillard *et al.*, 2009) that tend to be obscured by the increased presence of CMEs near solar maximum (Plotnikov *et al.*, 2016). The generation of the HICAT catalogue, and hence the current analysis, avoids significant contamination by such events by selecting only features of PA width greater than 20°. When considering the LASCO CMEs categorised as wide (> 20°) and "good" in Figure 3e, the daily rate of LASCO CMEs reduces to only 0.3 CMEs/day at solar minimum and 3 CMEs/per day at solar maximum. Note that the categorisation of a good LASCO event is derived only through interpretation of the contents of a "remarks" column in the CDAW catalogue in which the observer will comment on event quality, but without a strict definition. This is not only subjective, but will also undoubtedly vary from observer to observer, and possibly from day to day. For the purposes of the current study, we consider the good LASCO events to be those not referred to as "poor" or "very poor" in the remarks column (and poor events to be those specifically referred to as either poor or very poor; *i.e.* there is no classification of fair events). The fact that the daily rate of good, wide CDAW CMEs is highly consistent with the total COR-2 CME rates may imply that the majority of CMEs identified in COR-2 imagery are also wide, of course under the assumption that the COR-2 cataloguing philosophy focusses on events that would be deemed good by the CDAW cataloguers. Again, the daily rate of good, wide CDAW CMEs is not inconsistent, albeit somewhat higher, than the HI-1 CME rate, even accounting for differences in field of view extent. We draw the reader's attention to the work of Vourlidas *et al.* (2017), who performed a detailed comparison of COR-2 and LASCO CDAW CME rates over the period from 2007 to 2014; the authors come to similar conclusions in terms of the comparison consistency of COR-2 CME rates with LASCO CME rates when those listed as poor or very poor were excluded.

Given the uncertainties that arise from the fact that the different catalogues have been derived from different instruments, using different identification philosophies (including the subjectivity of manual selection), the calculated daily CME rates from HI-1, COR-2 and LASCO are reasonably consistent over the whole solar cycle. The fact that the rate of CMEs detected in the corona increases by such a small factor relative to the increase in sunspots from solar minimum to maximum has been known for many years (see Webb and Howard, 2012, and references therein), and we can now confirm that the same applies to CMEs in the heliosphere (Figure 3f). It is also interesting to note that, even at times of very few sunspots, in particular in late 2008 and early 2009, there were still significant numbers of CMEs detected in the corona and, correspondingly, the heliosphere.

Figure 4 presents the monthly CME rates for HI-1A (top left; all HICAT events), HI-1B (top right; all HICAT events), COR-2A (middle left), COR-2B (middle right) and LASCO (bottom left/right), each as a function of the monthly sunspot number. Linear best (least squares) fits are plotted in red on each panel; COR-2 panels include fits to the monthly rates of all CMEs (grey symbols, dotted red lines) and fits to monthly rates of the subset of CMEs identified as being wider than 20° (black symbols, solid red lines). Pearson correlation coefficients (marked on each panel) of $R$=0.89 (0.79) and 0.88 (0.66), respectively, indicate that the COR-2A and COR-2B coronal rates of all CMEs (wide CMEs) are well correlated with sunspot number. Corresponding $Y$-axis intercepts of +20.60 (+3.66) and +17.39 (+2.84) suggest, as noted earlier, the emergence of some CMEs even in the total absence of sunspot activity. HI-1A and HI-1B CME rates are similarly well correlated to sunspot number ($R$=0.79 for both), again with indication of a low level of CME emergence when no sunspots are present. Based on an examination of Figure 3, the slope of the best fit line to the sunspot number is expected to be around 4 or 5 times as large for all COR-2 CMEs than for HI-1 CMEs; this is in fact the case. The fact that the slope of the best fit line to wide (*i.e.* >20°) COR-2A CMEs is twice what it is for wide COR-2B events simply reflects the fact that more COR-2A CMEs have undergone further analysis to retrieve their observational and kinematic characteristics. This is most likely due, at least in part, to the lower signal-to-noise ratio of the synoptic COR-2B total brightness observations because of increased stray light background (see Vourlidas *et al.,* 2017). For completeness, the bottom panels of Figure 4 present monthly CME rates from LASCO, again derived from the CDAW event list, as a function of sunspot number; the left-hand panel includes all events with width >20° and the right-hand panel only good events (see previous discussion), again with widths > 20°. Note that, in the CDAW catalogue, around half of all events wider than 20° are listed as being poor or very poor, evidenced by best fit slopes of 1.22 and 0.55 for all and good events, respectively. The LASCO panels also confirm the presence of CMEs in the absence of sunspots.

The general increase in coronal and heliospheric CME rate with increasing sunspot number is clear, and, of course, unsurprising. This underlying trend is countered by a significant spread of CME rates for a specific sunspot number, demonstrating that, whilst there is a general trend for increasing numbers of CMEs with increasing solar activity, the CME rate for a specific level of activity can vary considerably.

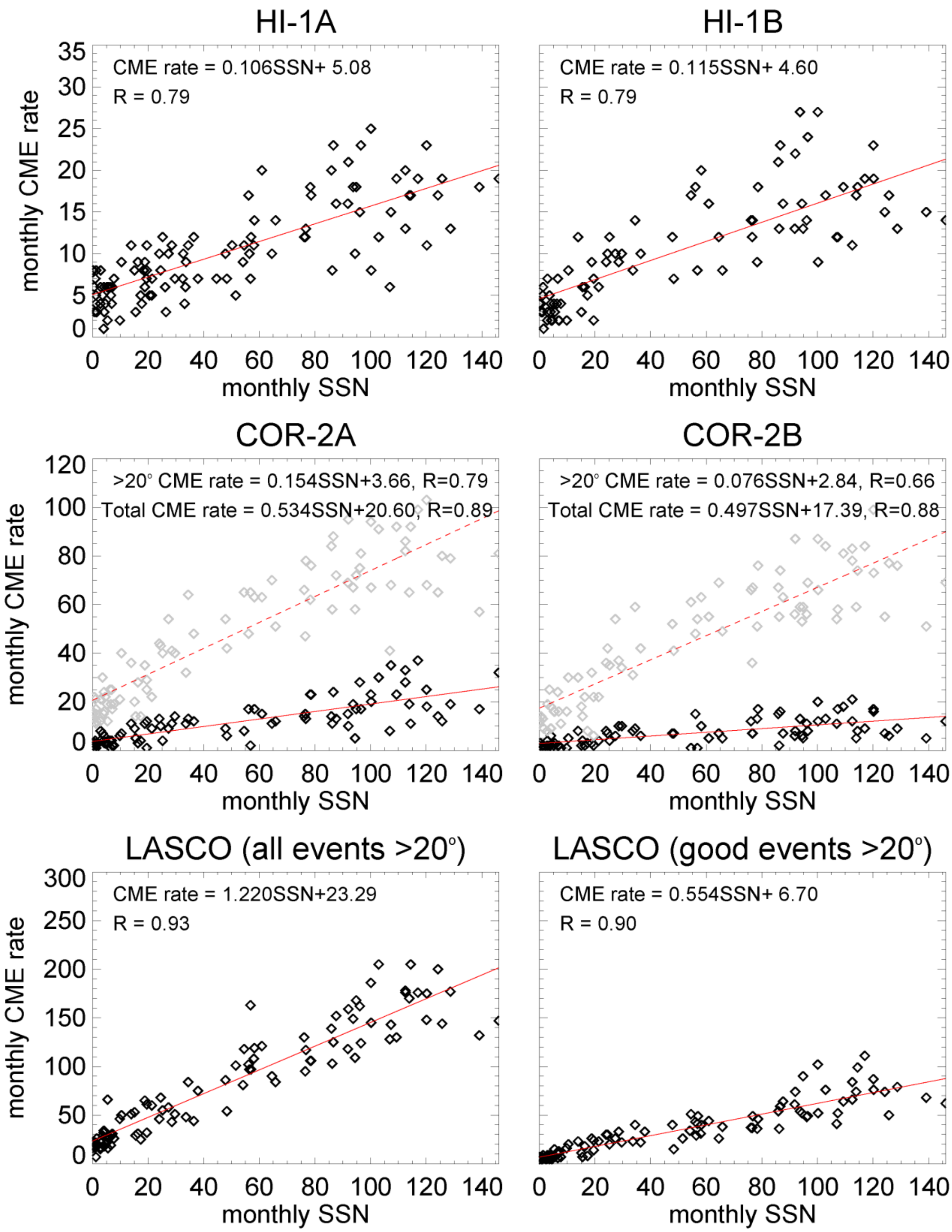

**Figure 4** Monthly CME rate as a function of monthly sunspot number for STEREO/HI-1A (top left; all events), STEREO/HI-1B (top right; all events), STEREO/COR-2A (middle left; all events, grey symbols, and events > $20^o$, black symbols), STEREO/COR-2B (middle right; all events, grey symbols, and events >$20^o$, black symbols), SOHO/LASCO (bottom left; all events >$20^o$, bottom right; good events >$20^o$). For each, red lines indicate least-square fits (the equation of the line, in the form $y = mx + c$, and the Pearson correlation coefficient is quoted on the corresponding panel).

The STEREO mission configuration, coupled with the viewing geometry of the HI instruments (see *e.g.* Eyles *et al.,* 2009), means that a CME may be detected by neither, either or both HI-1 instruments, depending critically on its propagation direction (including its angular width) and also on mission phase; as noted above, orbital evolution means that this likelihood of a given CME being observed by HI-1 on both spacecraft will vary markedly through the mission. To investigate this at a simplistic level, we compare the HICAT lists for STEREO-A and STEREO-B by simply asking the question: when we first detect a CME using the HI-1 instrument on one spacecraft, do we (first) observe a CME with the same instrument on the other spacecraft within a specified time window? This endeavour is a purely computational exercise; we make no attempt to ascertain through examination of the imagery whether we believe that it is the same CME; the presumption is that it is. We consider time windows of +/-1, 2, 3, 6, 9, 12 and 48 hours in duration. We undertake this comparison, initially, by identifying HI-1B CMEs that are first detected within these pre-defined time limits of the times at which HI-1A CMEs are first observed, and then vice versa. The resultant percentages of so-called "coincident events" are presented in the upper and lower panels, respectively, of Figure 5, as a function of time; data are binned using a bin size of three months. This figure includes events from April 2007 to August 2014; obviously no post-conjunction events are included due to the loss of contact with STEREO-B.

The time axis covers a range of separation angles between the two spacecraft. As an illustration, in January 2008, 2009, 2010, 2011, 2012, 2013, and 2014, the STEREO-A/Sun/STEREO-B separation angle ─ centred approximately on the Sun-Earth line ─ was 44$^o$, 88$^o$, 132$^o$, 176$^o$, 218$^o$, 260$^o$ and 302$^o$, respectively. At the start of the science phase of STEREO, in April 2007, the HI-1A and HI-1B fields of view had virtually no overlap. At that time, only a relatively longitudinally-extended near Earth-directed CME could theoretically impinge into the field of view of both cameras (but such halo CMEs are notoriously faint and such Earth-directed events were rare during that phase of the mission). As the two spacecraft drifted away from Earth, the region of overlap between the HI-1A and HI-1B fields of view progressively increased, maximizing when they were separated by 180$^o$ (in early 2011). At this time, HI-1A and HI-1B would both be able, theoretically, to observe near-equatorial CMEs expelled at longitudes between -90$^o$ and +90$^o$ of the Sun-Earth line. Obviously there are possible effects on the visibility of CMEs that propagate well beyond the Thomson sphere, which we discuss below but, potentially, virtually all CMEs detected by one HI-1 camera would be visible to the equivalent camera on the other STEREO spacecraft. Subsequently, as the spacecraft separated beyond 180$^o$ (defined as above), moving towards superior conjunction, the overlap between the HI-1A and HI-1B fields of view progressively reduced.

Hence, for a constant level of CME activity, one might naively expect the percentage of coincident events to increase from 0% in 2007 to 100% in 2011 and subsequently reduce to near 0% as conjunction was approached. This trend for an increase in the percentage of coincident events towards spacecraft opposition, and a reduction thereafter, is clearly evident in Figure 5 for all but the longest time window. An increase in the CME activity potentially complicates this as it could lead to a situation where an increasing number of events are coincident by chance. No coincident events are detected for the shortest (+/-1 hour) length window until the third quarter of 2008, after which time the percentage of coincident events increases to a maximum of around 50% in 2011; this is followed by a decline in the percentage of coincident events. This trend is mirrored in the windows of increasing length, albeit with increasing percentages. The propagation direction of a CME, coupled with its angular extent and its speed, as well as mission phase, will dictate the time difference between its entrance into the HI-1A and HI-1B fields of view. Rudimentary modelling (not shown) demonstrates that, for the majority of "typical" CMEs that impinge into both HI-1 fields of view, the time difference between their entry into the HI-1A and HI-1B fields of view would be less than 18 hrs. Hence it is likely that increasing the length of the time window much beyond +/-18 hours will only increase the number of chance coincidences. The likelihood of such chance coincidences is, of course, also determined by the CME rate (as will be discussed further below). The maximum rate of CME detections, by either HI-1A or HI-1B, is around 0.8 per day (or one every 19 hours). For time windows significantly shorter than this (for example +/-1 hour), we can be fairly confident ─ over all phases of the mission ─ that HI-1A or HI-1B are imaging the same CME.

The increase in CME rate towards solar maximum obviously complicates the interpretation of Figure 5. As noted above, for the shortest time window of +/-1 hour ─ which is much shorter than the average time between CMEs, even at solar maximum ─ we can be relatively confident that most identified coincidences are real (percentage coincidence peaks at around 40 to 50% in that case). However, the expected time difference between a CME's entry into the fields of view of HI-1A and HI-1B would generally be longer than an hour, which means that we are likely severely underestimating the percentage of coincident events with such a short time window. As the length of the window increases, the percentage of coincident events increases (for a window of +/-6 h, for example, the percentage coincidence maximises at 80%). However, the distribution also becomes increasingly asymmetric around its peak, with somewhat higher percentages post peak suggesting the increased incidence of chance coincidences that would be expected due to higher CME activity at this time. This is particularly evident for the longest time window, which ─ at +/-48 hours ─ is much longer than the average time of 19 hours between adjacent solar maximum CMEs.

That said, our principal conclusion is that the basic form of the distribution of coincidence events shown in Figure 5 is determined by events that are truly visible to both HI-1 cameras. We also note that the peak in percentage coincidence occurs before the period of maximum CME activity, supporting the validity of this underlying result. We should also mention, at this point, the influence of the Thomson scattering geometry on the observation of CMEs in the heliosphere, as addressed by such authors as Vourlidas and Howard (2006) and Howard and Tappin (2009). The latter introduced the idea of the Thomson plateau to describe the fact that plasma density features such as CMEs are observable over a large range of scattering angles and not just in the vicinity of the Thomson sphere, although their observability does drop off markedly as they propagate well beyond the Thomson sphere. Where we believe that the results of Figure 5 are mainly due to real CME coincidence (*i.e.* for windows of +/- 6 hours or shorter), we can argue that the broad peak in coincidence coupled with the relatively high percentage coincidence values (notwithstanding the issues discussed above) provide confirmation of the validity of this Thomson plateau approach, *i.e.* that the influence of Thomson scattering geometry on the observability of CMEs in the heliosphere is relatively small.

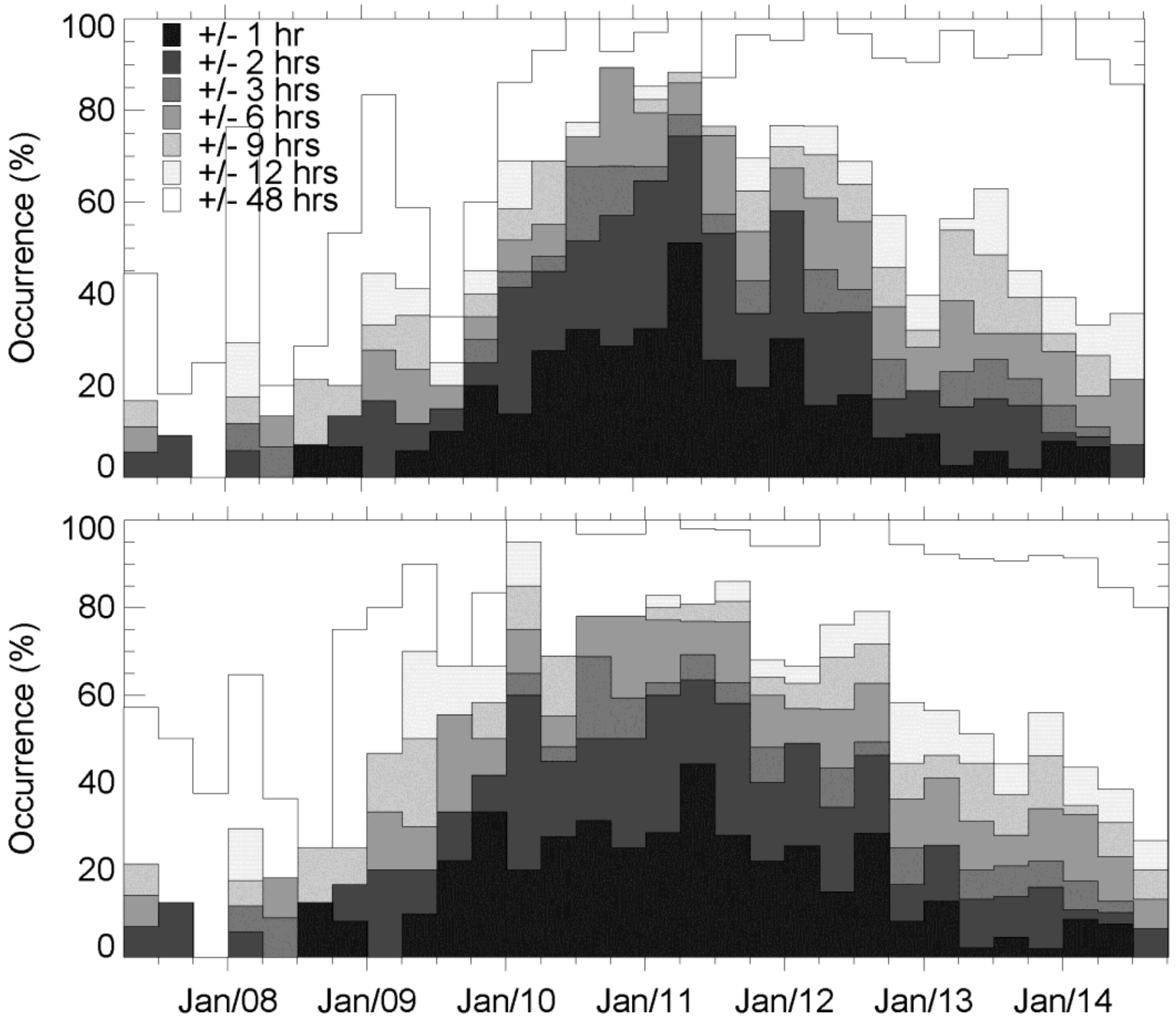

**Figure 5** Distribution of the percentage of so-called coincident CMEs, with a bin size of three months, for a range of different time windows (as noted in the upper panel). The top panel illustrates the percentage of CMEs identified in HI-1A imagery for which there is a coincident event identified by HI-1B. The bottom plot represents CMEs identified in HI-1B images for which there is a corresponding event seen by HI-1A. All HICAT CMEs (*i.e.* good, fair and poor) are included in this figure.

### 3.2 CME Position Angles and Widths

The top right pair of panels of Figure 6 present distributions of the central PAs of all heliospheric CMEs in the HICAT manual catalogue, stacked by year (with in-year normalization). The upper panel of the pair is based on imagery from HI-1A, the lower panel, on imagery from HI-1B. Corresponding distributions for coronal CMEs, observed by COR-2A, COR-2B and LASCO, are presented in the top right, bottom right, and bottom left pairs of panels, respectively. Prior to superior conjunction, the HI-1A and HI-1B instruments viewed (solar) east and west of the Sun-spacecraft line, respectively. Having emerged from superior conjunction in July 2015, the STEREO-A spacecraft was rolled by 180$^o$ to maintain communication with Earth; since that time, HI-1A views over the west solar limb (as viewed from the spacecraft). Hence HI-1A has a different scale on the bottom and top X-axes, pertaining to east (0$^o$ – 180$^o$) and west (180$^o$ – 360$^o$) limb observations. As noted above, the range of possible central PAs for HI-1 CMEs is limited by geometry of the HI field of view. The top panel of each pair of COR-2A, COR-2B and LASCO panels corresponds to CMEs with central PAs over the east limb (*i.e.* between 0$^o$ and 180$^o$); the bottom panel corresponds to west limb CMEs (central PA between 180$^o$ and 360$^o$). Note that for LASCO, only good CMEs from the CDAW catalogue that are wider than 20$^o$ are included in the analysis; for COR-2, only the relatively small subset of analysed CMEs that have derived widths greater than 20$^o$ are included (we assume that these CMEs are representative of the full catalogue of events). Diamonds and crosses denote the median and quartile values of central PA for each year.

During the solar minimum years of 2008 to 2009, the central PA of coronal and heliospheric CMEs was generally confined to PAs close to 90$^o$ and 270$^o$ for east and west limbs events, respectively (and hence, by inference, to more equatorial latitudes). Most solar minimum CMEs are limited, in terms of their central PA, to within ≈ 30$^o$ of the solar equator. From 2010 to, say, 2014, as solar activity increases towards, and enters, the solar maximum period, the distribution of central CME PAs widens across the entire range of PAs accessible to COR-2 and LASCO. Although the corresponding central PA distribution for heliospheric CMEs does broaden, it does so to a lesser degree, appearing to remain more concentrated around the equator than are the coronal CMEs during this time. Thus, although potential CME-generating regions are likely to be more widely distributed in latitude when the Sun is more active, outward propagating CMEs appear to be deflected equatorward despite the more complex nature of the coronal structure at such times (this will be discussed in more detail below). Over the declining years of Solar Cycle 24 (after 2014), we are limited to data from HI-1A and LASCO only; during this time, the central PA distribution of CMEs identified in LASCO imagery remains broad, whereas the corresponding distribution for HI-1A becomes narrower. We note that the CDAW and COR-2 catalogues both list CME central PAs explicitly, as opposed to HI which just cites the northermost and southernmost extent from which the central PA is derived. The CDAW catalogue identifies full halos (*i.e.* those encompassing the full 360$^o$ PA extent of the coronagraph field of view) for which it does not provide a central PA; hence they are not included in the generation of Figure 6. Full halo events that are identified as such within the COR-2 catalogue are also excluded in the generation of Figure 6.

Hence, the tendency for CMEs to "emerge" over a wider range of latitudes (both northern and southern) at solar maximum although evident, to some degree, for heliospheric CMEs is much clearer for coronal CMEs; we note that for both coronal and heliospheric CMEs, the median central PA value remains close to equatorial throughout indicating no significant hemispheric bias. In fact, the heliospheric CMEs show remarkably consistent median values, near to the equator compared to the much greater variation of the coronal median values. Unlike the migration of sunspots to latitudinally-confined northern and southern bands with increasing solar activity, evidenced by the well-known butterfly diagram, the distribution of central CME PA in both the corona and heliosphere simply shows an expansion over a wider range, but with the bulk of events still being confined to lower PAs. The fact that CMEs are observed over a wider range of PAs (and thereby latitudes) at solar maximum confirms the results of a plethora of previous studies of coronal CMEs (e.g. Yashiro *et al.* 2004). This is strong evidence that, although CMEs are associated with active regions, the simple-minded view that CMEs originate from active regions alone, or are centred on active regions, cannot be readily supported; such a scenario ought to reveal a poleward migration of CME central PAs in keeping with the sunspot migration. In other words, the results do not support the so-called standard model in which CMEs generally lie directly above flares/active regions (see *e.g.* Harrison, 1995).

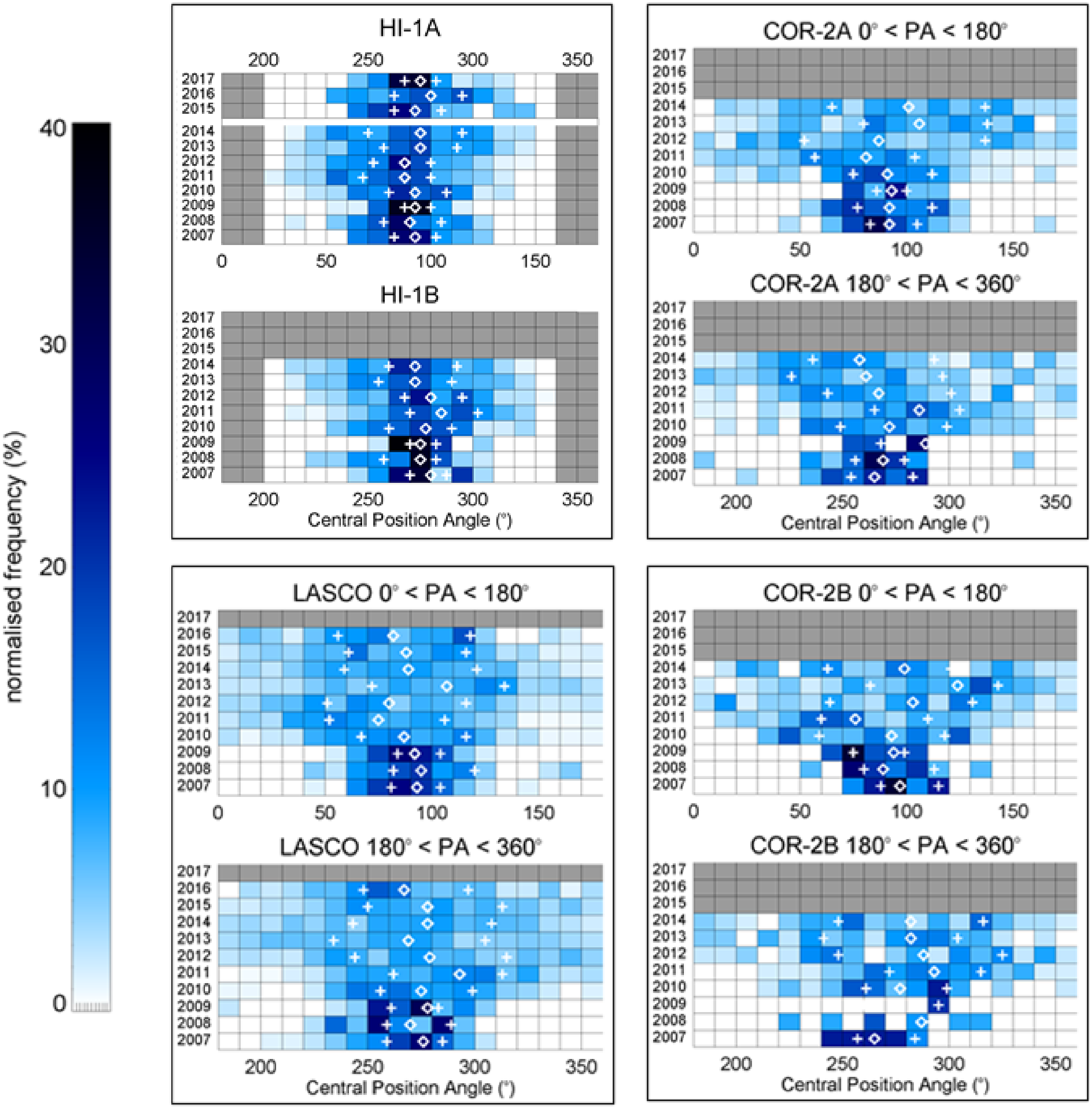

**Figure 6** Stacked annual distributions of CME central PA (in 10$^{o}$ bins) for HI-1A (upper panel of top right pair), HI-1B (lower panel of top right pair), COR-2A (top left), LASCO (bottom left) and COR-2B (bottom right). To aid comparison, for each coronagraph, the upper and lower panels of each pair correspond to events detected over the east and west limbs, respectively. Each annual distribution is normalised such that it adds up to 100%. Diamonds and crosses denote the median and quartile values of central PA, again for each year. In the case of LASCO, only good CMEs with PA widths > 20$^{o}$ are included. For COR-2 only CMEs with derived PA widths > 20$^{o}$ are included. For HI, all (*i.e.* good, fair and poor) events are included (HICAT CMEs are all wider than 20$^{o}$ in PA extent due to the cataloguing philosophy adopted).

The suggestion in Figure 6 that heliospheric CMEs tend to be much more confined to the equatorial zones than their coronal counterparts may imply equatorward drag on CMEs as they propagate from coronal to heliospheric regimes, due to the interaction of the CMEs with the global magnetic field. There has long been evidence for such equatorward drag on CMEs as they propagate into and through the corona. For example, Hildner (1977) and MacQueen *et al.* (1986) found PA (and hence inferred latitudinal) deflections of order 2.5$^{o}$ and 2.2$^{o}$, respectively, in observations of the first few solar radii above the Skylab coronagraph occulting disc. However, such a modest deflection is akin to what one might expect to be the accuracy in discerning the PA of typical CME structures, simply due to the diffuse nature of CMEs and their boundaries. Rather than comparing PA deviations in coronagraph image sequences, many recent studies have addressed the issue of latitudinal deflection of CMEs by comparing their coronal signatures with perceived source regions (*e.g.* Cremades and Bothmer, 2004; Wang *et al.,* 2011; Isavnin *et al.*, 2014; Liewer *et al.*, 2015; Kay *et al.*, 2017) although such studies inherently include assumptions about the nature of the relationship between the visible-light coronal CME structure and the source region, the latter most commonly observed in EUV (see *e.g.* Harrison *et al.*, 2012, and references therein). Kay *et al.* (2017), in a study of this type, found deflections for seven CMEs in the range 6.8° to 23.3° with respect to a polarity inversion line (PIL) source region. With the advent of heliospheric imaging, one can now compare PAs/latitudes of CMEs in the corona and heliosphere for greater clarification of such deflection, without the need to make assumptions about source locations. Some studies involving coronal and heliospheric observations have shed some light on this issue, providing evidence for deflection of individual events (*e.g.* Byrne *et al.*, 2010; Harrison *et al.*, 2012).

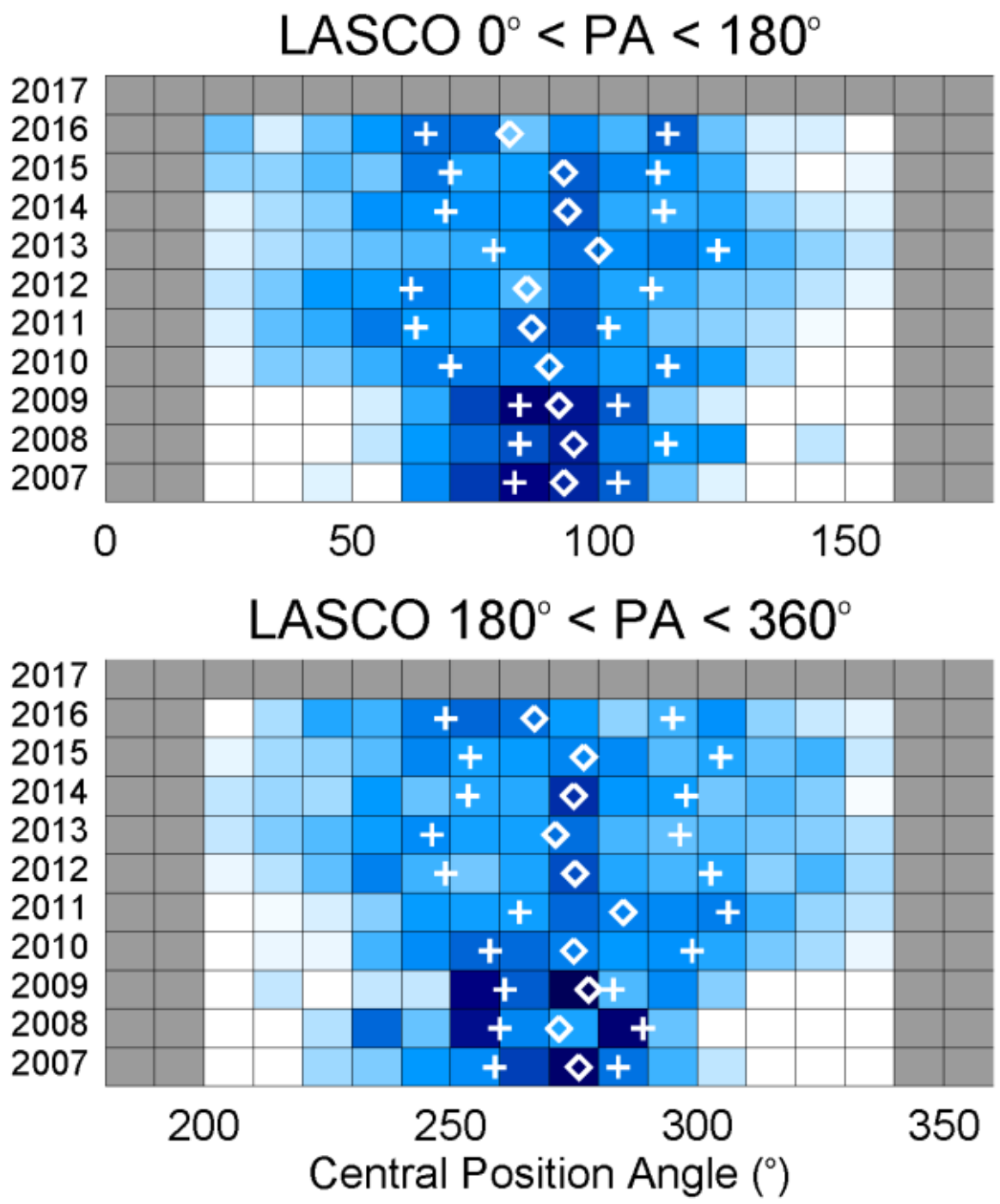

Figure 7 A reproduction of the LASCO panels of Figure 6, but with the central PA of each CME calculated under the assumption that the northernmost and southernmost PA of each CME are, if applicable, truncated to the corresponding limit of the HI-1 field of view.

We note that the HI-1A and B panels of Figure 6 (top left) are based on the analysis of all HICAT CMEs, including those that appear to extend beyond (northward, southward, or both northward and southward of) the HI-1 field of view. The catalogue entry for the northern or southernmost extent of such an event is truncated, by necessity, to the corresponding limit of the field of view (although with an identifier to indicate that fact). As was mentioned previously, the central PA of a HICAT CME is calculated as the PA midway between the northernmost and southernmost PA extents of the CME. In the case where a CME actually extends beyond the field of view, this method of determining its central PA will generally lead to a value that is artificially equatorward of its true central PA. In an attempt to assess the potential impact of this bias on the HI-1 distribution, Figure 7 reproduces the LASCO panels of Figure 6 but with the central PA of each CME calculated under the assumption that the northernmost and southernmost PA of each CME are, if applicable, truncated to the corresponding limit of the HI-1 field of view. Note that, for the purposes of this analysis, we neglect the fact that the northernmost and southernmost PA limit of the HI-1 field of view varies by +/-7.25° over the year, as the boresight of the HI fields of view are aligned on the ecliptic rather than the solar equator. We tentatively deduce, from comparison of the LASCO panels of Figure 6 with those in Figure 7, that this instrumental effect would not be solely responsible for the restriction of the distribution of the central PA of HI-1 CMEs to equatorial latitudes. This may support, statistically, the idea of continued equatorward drag of CMEs into the heliosphere.

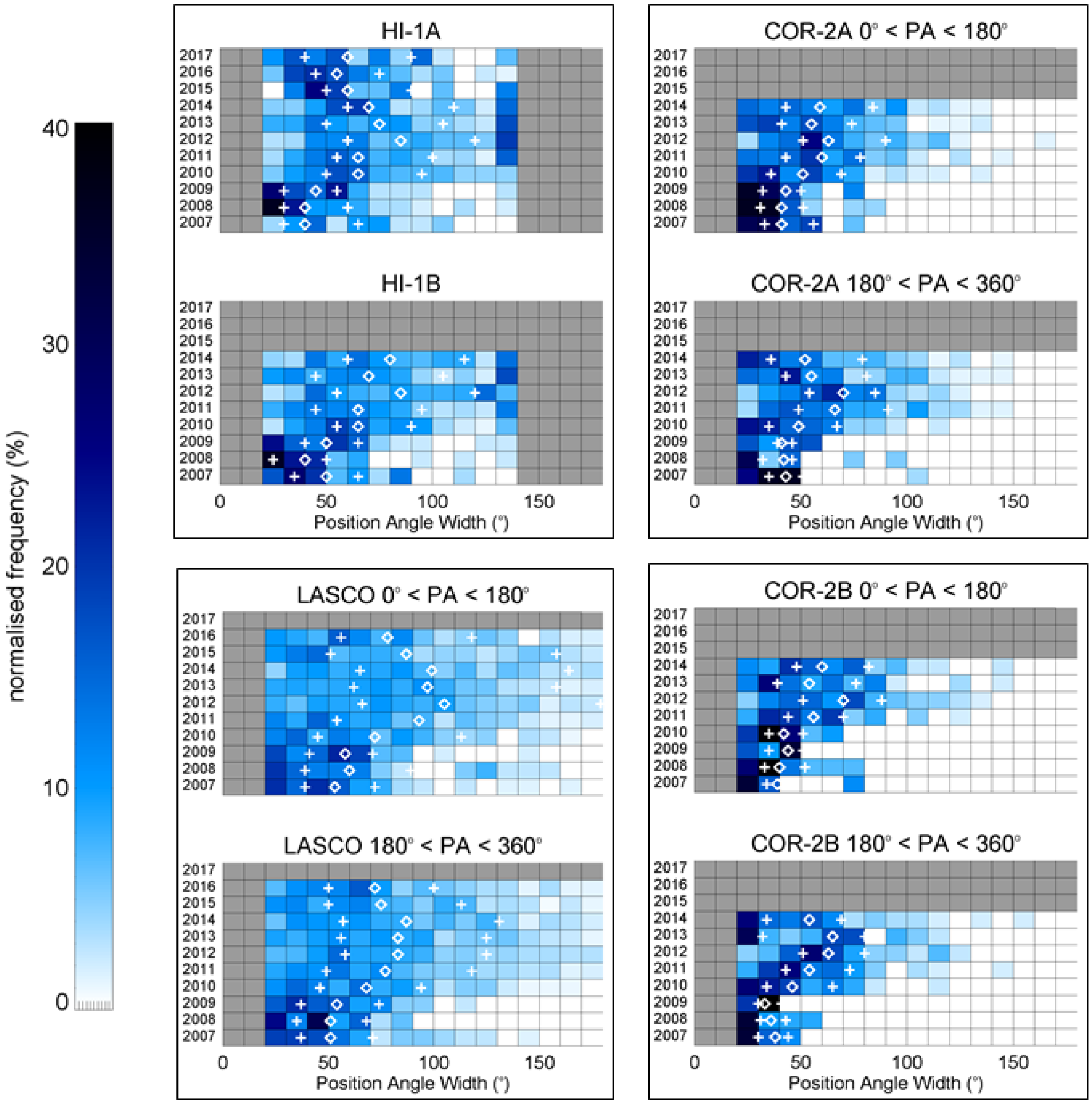

**Figure 8** Stacked annual histograms of CME PA widths (in $10^{o}$ bins) for HI-1A (upper panel of top right pair), HI-1B (lower panel of top right pair), COR-2A (top left), LASCO (bottom left) and COR-2B (bottom right). To aid comparison, for each coronagraph, the upper and lower panels of each pair correspond to events where the central PA of the event is over the east and west limb, respectively. Each annual distribution is normalised such that it adds up to 100%. Diamonds and crosses denote the median and quartile values of PA width, again for each year. The figure is based on the same CMEs as used in the generation of Figure 6.

Figure 8 presents analogous distributions of PA width for coronal and heliospheric CMEs, again illustrated as a function of year with in-year normalisation. As in Figure 6 (and 7), upper and lower panels of the pair based on COR-2 and CDAW CME catalogues correspond to events in which the central PAs is over the east limb (*i.e.* between $0^{o}$ and $180^{o}$) and west limb CMEs (i.e. between $180^{o}$ and $360^{o}$), respectively. For HI-1, upper and lower panels are based on CME identifications in HI-1A and HI-1B imagery, respectively; so, although the HI-1B panel includes only west limb CMEs, HI-1A includes CMEs centred over the east limb prior to (and including) 2014 and west limb thereafter. For each instrument, the CMEs used in the generation of the figure are the same as those used to create Figure 6. It is clear that the PA width of a CME, imaged in both the corona and heliosphere tends to be narrower during solar minimum years (2008-2009) than during years of higher activity (2011-2014). The post-conjunction years (2015 to 2017) show a corresponding reduction in width as the activity declines. That is not to say that narrower CMEs are absent around solar maximum. The median heliospheric CME is typically 40-$50^{o}$ wide in PA during quiet solar conditions, increasing to 70-90$^o$ around solar maximum. The same general increase in CME width with increasing solar activity is reflected in the coronal (COR-2 and LASCO) data; however it is difficult to compare coronal and heliospheric PA widths since the LASCO CDAW CMEs are listed consistently as being wider than those from COR-2 (see also, Vourlidas *et al.,* 2017, who noted this but did not explain the discrepancy). We do need to note that the COR-2 CMEs that have been fitted to yield the observational characteristics of PA width and central PA are a relatively small – and possibly biased – subset of the whole.

Yashiro *et al.* (2004) reported on the variation of the widths of LASCO CMEs in the CDAW catalogues over the previous cycle (Solar Cycle 23) and, as mentioned previously, noted an increase of median PA width of only around 20$^o$ from minimum to maximum; this is a smaller increase than that indicated by the corresponding panels of Figure 8. A number of the wider heliospheric CMEs during the solar maximum years (**2012**-2014), in particular, are artificially truncated in PA width due to the limited coverage of the HI-1 field of view, leading to spikes in event occurrence in the right hand column of the HI-1A and HI-1B distributions in Figure 8.

We interpret the results of Figure 8 as confirming that CMEs in the heliosphere, like those in the corona, tend to be narrower (median PA width 40-50$^o$) at solar minimum than at solar maximum (median PA width up to 80$^o$). This result is based on observations from Solar cycle 24; how we reconcile that with the Solar Cycle 23 results reported by Yashiro *et al.* (2004) - which indicate that the PA spans of CMEs at least in the corona were rather less clearly dependent on the phase of the cycle - is open for debate.

## 4. Discussion and Conclusions

In this paper, we present the first truly statistical analysis of the observational characteristics of CMEs imaged in the heliosphere by the HI-1 instruments on STEREO, based on the manually-generated HICAT catalogue created under the auspices of the EU FP7 HELCATS project. In that sense, this paper provides a benchmark for future heliospheric CME analyses. Comparisons of heliospheric and coronal properties of CMEs are crucial for understanding, for example, how such events evolve as they propagate through the solar system, for understanding stellar mass-loss processes, and for assessing the nature of the heliosphere and impacts on solar system bodies.

The principal conclusions of this work are the following:

i) Unsurprisingly, the general trend in heliospheric CME activity is consistent with the solar activity cycle, increasing from a minimum in the years 2008-2009 to a maximum in 2012-2013, followed by a subsequent decline to 2017. Step-wise increases in the rate of heliospheric CMEs occur near the start of 2010 and 2011.

ii) CME rates estimated for the whole heliosphere for all event categories (good, fair and poor) included in the HICAT catalogue are around 0.3-0.4 CMEs per day and ≈ 1.4-1.6 CMEs per day for the last solar minimum (2008-2009) and maximum (2012-2014), respectively; these are not inconsistent with the coronal CME rates recorded from the STEREO and SOHO spacecraft, once differences in cataloguing philosophy are considered. The deep minimum between Solar Cycles 23 and 24, despite including long periods of no sunspots, was associated with the ejection of significant numbers of CMEs, identified at both coronal and heliospheric altitudes.

iii) The orbital configuration of STEREO is such that the degree to which the same CMEs are imaged by the HI-1 cameras on both spacecraft depends critically on mission phase, with few CMEs being observed from both vantage points during the early years of the mission rising to between 40% and 90% of CMEs being observed by both HI-1 cameras; the percentage of such so-called coincident events depends on the length of the time-window used. This peak coincides, as expected, to that phase of the mission where the spacecraft were in opposition (*i.e.* separated by 180$^o$) at which time, early in 2011, the overlap of the HI-1A and HI-1B fields of view maximises. The percentage of coincident events reduced thereafter. We suggest that the results confirm previous assertions that suggest that Thomson scattering geometry is not a critical factor in imaging CMEs.

iv) We find that, during solar minimum years, CMEs in the heliosphere are more restricted to equatorial latitudes than during solar maximum years. Although the median value of the central PA of heliospheric

CMEs remains largely within 10$^{o}$ of the equator over the entire solar cycle, its distribution extends over a much broader range of PAs (hence, by interference, latitudes) with increasing activity. This is consistent with the behaviour of coronal CMEs revealed by this and previous studies.

v) This provides strong evidence to support the view that, although CMEs are associated with active regions, a simplistic view that CMEs originate from active regions alone, or are centred on active regions, cannot be readily supported; we find a broadening of the distribution of central CME PAs as we move towards solar maximum, rather than a migration to higher values. As in the corona, CMEs in the heliosphere do not obey an equivalent to the Spörer sunspot law and, whilst the CME association with active regions is not questioned, it is clear that a more complex source model is required to account for this behaviour.

vi) A more detailed comparison of the distribution of the central PA of heliospheric and coronal CMEs provides tentative evidence for the equatorward drag of CMEs as they propagate outwards from the Sun, in particular around solar maximum. This is an important extension of earlier studies of the latitudinal deflection of CMEs.

vii) The distribution of the PA width of heliospheric CMEs mirrors the well-known increase in the width of coronal CMEs towards solar maximum, with the median width of CMEs in the heliosphere increasing from 40-50$^{o}$ in PA at solar minimum to a value nearer 80$^{o}$ at maximum. However, it is important to note that the observations show that a CME of any apparent width can be observed at any time. While this increase in CME PA widths as a function of solar cycle is reflected in the coronal CMEs over the same period, it is noted that such a broadening with solar activity was not so clear for the previous solar cycle (Yashiro *et al.,* 2004). Moreover, there is some evidence that CMEs tend to be "clearer" (brighter and more well-formed) at solar maximum.

As expected, many of the conclusions of this work confirm the results of previous analyses of the statistical behaviour of CMEs in the corona, extending these conclusions to the case of the heliosphere. This includes results pertaining to CME rates, the variation of CME characteristics with solar cycle, and evidence supporting the continued deflection of CMEs towards the equator as they propagate outward, although, unlike some previous studies, the HI-1 data indicate such deflection predominantly occurs during solar maximum.

The HICAT catalogue, on which this analysis is based, provides a benchmark for the analysis of heliospheric CMEs that sits alongside established coronal CME catalogues, enabling for a more complete view of CMEs from source/onset studies to impacts at Earth and elsewhere. Note that an analysis of the kinematic properties of many of the HICAT CMEs forms the basis of a follow-on paper (Barnes *et al.,* 2018).

Finally, we remind readers that these and other event lists derived as part of the HELCAT projects can be accessed from the HELCATS website at www.helcats-fp7.eu.

**Acknowledgements** We acknowledge support from the European Union FP7-SPACE-2013-1 programme for the HELCATS project (#606692). The HI instruments on STEREO were developed by a consortium that comprised the Rutherford Appleton Laboratory and the University of Birmingham (both in the UK), Centre Spatial de Liège (CSL), Belgium, and the Naval Research Laboratory (NRL), USA. The STEREO/SECCHI project, of which HI is a part, is an international consortium led by NRL. We recognise the support of the UK Space Agency for funding STEREO/HI operations in the UK. We also acknowledge use of the CDAW SOHO/LASCO event catalogue (https://cdaw.gsfc.nasa.gov/CME_list/), which is generated and maintained at the CDAW Data Center, by NASA and the Catholic University of America in cooperation with NRL, and the STEREO/SECCHI/COR2 CME catalogue (http://solar.jhuapl.edu/Data-Products/COR-CME-Catalog.php) that is generated and maintained at Johns Hopkins University Applied Physics Laboratory, in collaboration with NRL and the NASA Goddard Space Flight Center. We thank the Royal Observatory, Belgium for provision of sunspot number data.

**Disclosure of Potential Conflicts of Interest** The authors declare that they have no conflicts of interest.